# Residual-resistance-ratio of Cu stabilizer in commercial REBCO tapes

*Jun Lu, Yan Xin, Vince Toplosky, Jeremy Levitan, Ke Han, Jane Wadhams, Munir Humayun, Dmytro Abraimov, and Hongyu Bai*

National High Magnetic Field Laboratory, Tallahassee, FL, USA

*Yifei Zhang,*

SuperPower Inc. Glenville, NY, USA

**Abstract**

Residual-resistance-ratio (RRR) of Cu stabilizer in REBCO coated conductor is an important design parameter for REBCO magnets. In this work, we measured RRR of electroplated Cu stabilizer in commercial REBCO tapes. Over 130 samples were measured for the quality assurance programs of REBCO magnet projects at the National High Magnetic Field Laboratory, USA (NHMFL). The average RRR value was above 50. In order to investigate the factors that influence RRR, several samples were analyzed by using scanning electron microscopy, secondary ion mass spectroscopy, and inductively coupled plasma mass spectroscopy. We found that, in our samples, RRR was strongly correlated with the grain size. We demonstrated that RRR was primarily determined by grain boundary resistivity. Lower RRR was also strongly correlated with higher concentration of chlorine impurity. This is explained by that higher chlorine impurity hindered the grain growth in the room temperature self-annealing process resulting smaller grain. Smaller grain resulted in lower RRR. In addition, thermal annealing significantly enhanced RRR. An activation energy of 0.4 eV was obtained from the annealing experiment which corresponds to the activation of Cu grain growth.

**Key words**: Cu, residual-resistance-ratio, superconductor, REBCO

1. Introduction

In a superconducting composite wire, a certain amount of stabilizer material with high electrical and thermal conductivity is necessary. Electrical conductivity of a metal at low temperatures is often assessed by residual-resistivity-ratio (RRR), defined as a ratio of resistivity at room temperature and that at 4.2 K. The higher the RRR, the higher the electrical conductivity and the thermal conductivity at low temperatures, as thermal conductivity of a metal is proportional to electrical conductivity. RRR of the stabilizer is a key design parameter for superconducting magnets.

High RRR Cu stabilizer in low temperature superconductor (LTS) wires is essential for stability of LTS magnets [1]-[3]. For this reason, specification for RRR was developed for commercial $Nb_3Sn$ wires used in large LTS magnet systems [4]-[6]. For REBCO magnets, thermal instability does not seem to be a big concern thanks to fact that the critical current of REBCO is less sensitive to temperature than LTS. Nevertheless, RRR value of Cu stabilizer in REBCO tapes can still significantly influence several aspects of magnet operation, such as quench detection and protection. Although a minimum RRR requirement for a specific REBCO magnet needs detailed quench calculations, a high RRR is always desirable from the point of view of quench protection. The Cu stabilizer in most commercial REBCO conductors is electroplated and has RRR values between 20 and 60 [7], which are significantly lower than RRR of Cu in NbTi and $Nb_3Sn$ wires. National High Magnetic Field Laboratory have developed a 32 T all-superconducting user magnet [8] using REBCO coated conductor. A 40 T REBCO magnet is currently being developed [9]. For these two projects, RRR of the Cu stabilizer of the REBCO conductor was specified based on the historical values achievable by the conductor manufacturer.

RRR of Cu is usually measured in zero magnetic field. In high magnetic fields and at low temperatures, the magnetoresistivity becomes very important. Therefore, it is prudent to understand the effect of magnetoresistivity on Cu conductivity. According to Matthiessen's rule, resistivity ρ of a pure metal may be written in the form,

$$\rho = \rho_{thermal} + \rho_{impurity} + \rho_{defect} + \rho_{magneto} \qquad (1)$$

where $\rho_{thermal}$ is the resistivity due to thermal vibration of the lattice which leads to electron-phonon scattering. $\rho_{impurity}$ is due to the scattering by impurity atoms; and $\rho_{defect}$ is due to the scattering by structural defects such as dislocations and grain boundaries. The last term $\rho_{magneto}$ is the contribution from magnetoresistivity.

The term $\rho_{thermal}$ can be formulated by a Block-Gruneisen function and is dominant at room temperature even in high magnet fields. At temperatures below 20 K, However, $\rho_{thermal}$ is negligibly small. At these temperatures, $\rho_{impurity}$ and $\rho_{defect}$ are more important terms. Therefore, to reduce the resistivity at low temperatures, one needs to minimize the chemical impurities and structural defects. In high magnetic fields, magnetoresistivity becomes appreciable. When magnetic field is perpendicular to the electric current (transverse field), which is the case for most magnet applications, the magnetoresistivity $\rho_{magneto}$ in Cu at temperature $T$ can be described by the formula [10],

$$\mathrm{Log}(\rho_{magneto}/\rho_0(T)) = -2.662 + 0.3168 \log(B \cdot S(T)) + 0.6229 (\log(B \cdot S(T))^2 - 0.1839 (\log(B \cdot S(T))^3$$
$$+ 0.01827(\log(B \cdot S(T))^4 \qquad (2)$$

where $\rho_0(T)$ is the zero-field resistivity in Ω-m, $B$ is the magnetic field in tesla, $S(T) = \rho_0(273\ K)/\rho_0(T)$.

Equation (2) was used to calculate $\rho_{magneto}(B, 4.2\ K)$ for different $\rho_0(4.2\ K)$. Subsequently the effective RRR, which is defined as $\rho_0(295\ K)/(\rho_0(4.2\ K) + \rho_{magneto}(B, 4.2\ K))$, was calculated. Fig. 1(a) plots the effective RRR as a function of magnetic field for various zero-field RRR (RRR(0)) which corresponds to various $\rho_0(4.2\ K)$. For high RRR(0) materials, the effective RRR decreases quickly with magnetic field $B$. For instance, for RRR(0) = 150 the effect RRR is lowered to less than 30 in a 12 T field. At higher fields as shown in the inset, the magnetoresistivity makes such a prominent contribution that the effective RRR is lower than 20 for all RRR(0) values. The difference between RRR(0) = 150 and 50 is only 6% at 40 T. The benefit of high conductivity Cu in zero-field is apparently compromised in high fields. It is important to note, however, even at high fields, very low RRR(0) still has significant negative impact to the effective RRR. For example at 40 T, effective RRR of Cu with RRR(0) = 50 is still 70% higher than that with RRR(0) = 10.

In the case of the magnetic field parallel to the direction of electrical or thermal conduction (longitudinal field), such as the thermal conduction in axial direction of REBCO pancake coils, the magnetoresistivity is significantly smaller than that in the transverse direction and saturate at 3 - 10 T at 4.2 K [11]. If the magnet cooling is relied on thermal conduction in the axial direction, the effect of magnetic field is much smaller, and high RRR(0) should still be very important to thermal stability of the REBCO magnet.

In spite of the significant magnetoresistivity effect, RRR of the Cu stabilizer in REBCO conductor is still considered an important design parameter for high field REBCO magnets. So in this paper, we measured RRR of many samples taken from over 20 km of REBCO tapes manufactured by SuperPower Inc. Several samples were studied with a goal of achieving higher RRR values. Inductively Coupled Plasma Mass Spectroscopy (ICP-MS) and Secondary Ion Mass Spectroscopy (SIMS) were performed to measure the impurities in the Cu. Scanning electron microscopy (SEM) was used to investigate microstructure of these samples. The chemical impurity and microstructure were correlated to their RRR values. The effect of annealing on RRR was also investigated.

2. Experimental method

The samples are 4 mm wide SCS4050-AP REBCO tapes made by SuperPower Inc. The REBCO conductors for the 32 T magnet project have 50 μm Cu stabilizer electroplated by an external vendor; those for the 40 T magnet project have 20 μm Cu electroplated by SuperPower. In addition, several R&D samples dedicated to RRR studies were electroplated by SuperPower as well.

For RRR measurement, a 100 mm long sample was cut from each spool of conductor which was typically 100 - 200 m long. The surrounding Cu edges were trimmed by using a pair of scissors. Then the Cu/Ag stabilizer layer was peeled manually from the conductor as depicted in Fig. 2. The residual REBCO remained on the Ag layer was chemically removed by using a 0.6% of nitric acid solution for about 3 minutes. The peeled 100 mm long film that contains the 20 or 50 μm Cu and about 2 μm Ag was used for resistance measurement.

Four-probe resistance measurements were performed at room temperature (295 K) and in liquid helium in zero magnetic field. A pair of voltage taps typically 70 mm apart was soldered using $Pb_{37}Sn_{63}$ or $In_{49}Sn_{51}$ solders. Effort was made to reduce the size of solder spots to minimize the small positive error in the measured RRR due to superconductivity of $Pb_{37}Sn_{63}$ or $In_{49}Sn_{51}$ at 4.2 K. DC current of 1 A was delivered by a HP 6631B 10 V-8 A power supply. The voltage was measured by a Keithley 2010 digital multimeter.

Chemical trace element analyses were performed by Inductively Coupled Plasma Mass Spectrometry (ICP-MS). Cu/Ag film sample was dissolved by 14 N $HNO_3$ then acidified to form a 2 % $HNO_3$ solution that was directly analyzed. Elemental abundances were determined by ICP-MS using a Thermo Element XR™ equipped with an Elemental Scientific Inc. PFA spray chamber. The ICP-MS was tuned to yield > 1 million counts/sec on 1 part per billion (ppb) tuning solution of $^{115}$In. Peaks for 60 elements were monitored in low resolution mode. Concentrations in solution were converted into concentrations in the metal using gravimetrically determined sample weights.

Secondary Ion Mass Spectrometry (SIMS) was performed at Eurofins/EAG Laboratories to measure the depth profiles of O, Cl, P, S, and Fe in Cu. The depth profiling of Cu/Ag film began from the Ag layer with a 100 μm x 100 μm raster area. Cesium ion beam ($Cs^+$) was used for O, Cl, P, S analyses; and oxygen ion beam ($O_2^+$) was used for Fe analysis.

For microstructure analysis, Ga$^+$ ion beam imaging was performed using 24 pA current in a Thermal Scientific Helios G4 UC dual-beam field-emission SEM.

## 3. Experimental results

### 3.1 RRR of Cu stabilizer

RRR of 50 μm Cu stabilizer of REBCO tapes for the 32 T magnet projects are plotted in Fig. 3(a). The average RRR of the total 89 samples is 57 with a standard deviation of 4. The maximum and minimum RRR are 68 and 50 respectively. RRR of 20 μm Cu stabilizer of REBCO tapes for the 40 T magnet project are plotted in Fig. 3(b). The average RRR is 50 with a standard deviation of 9. The maximum and minimum RRR are 66 and 31 respectively. It was noticed that after initial 11 samples of RRR of about 60, there was a drop in RRR. Therefore, we started to investigate the factors that influence RRR.

### 3.2 Chemical impurities in Cu stabilizer

Chemical impurities, especially oxygen, are usually the main cause of low RRR in bulk Cu. We measured impurity concentrations of two samples with RRR = 25 and 60 respectively by ICP-MS. Valid data were obtained for more than 40 elements (not including O or Cl). No impurity were found above 1 ppm, except for Gd, Y and Ba which are evidently from the residuals of the REBCO layer.

SIMS depth profiles of O, P, S, Cl and Fe were performed on 3 samples with RRR = 25, 37, and 60 respectively. Fig. 4(a) shows their O depth profiles. The depth profiling started from the side of the Ag layer. High levels of O in depths 0 – 2 μm are due to the absorbed O in the Ag layer which is known to be permeable by oxygen. The O concentration quickly decreases and levels off in the Cu layer. The sample with the lowest RRR indeed has the highest O concentration. In addition, Cl depth profiles in Fig. 4(b) indicate that RRR values are strongly correlated with the Cl concentrations.

The measured impurities by ICP-MS and SIMS are summarized in Table 1 in ppm by weight. The data for O, Cl, and S are averages over depth of 6 – 10 μm from SIMS. Clearly O and Cl are the most prominent impurities. Despite the fact that O in Cu significantly reduces conductivity, the measured RRR values cannot be explained by the measured O alone. Because, for example, 7.2 ppm of O would result in RRR = 114 as estimated from ref. [10], more than 4 times higher than the measured value of RRR = 25. The direct effect of Cl on conductivity of Cu is not particularly significant [12]. The indirect impact of Cl on Cu resistivity may be significant, however, as will be discussed later. Above chemical analyses suggest that O, Cl and other impurities are not directly responsible for some low RRR values.

Table 1. Impurity concentrations in Cu stabilizers

| Element | Concentration (ppm in weight) | | |
|---|---|---|---|
| | RRR = 25 | RRR = 37 | RRR = 60 |
| O | 7.2 | 1.8 | 2.7 |
| Cl | 14.2 | 3.1 | 0.9 |
| S | 0.1 | < DL | < DL |
| P and Fe | < DL | < DL | < DL |
| Other elements | < 1.0 | - | < 1.0 |

### 3.3 Microstructures of Cu stabilizer

A significant contribution to resistivity comes from the electron scattering by structural defects, as represented by the term $\rho_{defect}$ in equation (1). Resistivity at grain boundaries, in particular, is very important [13]-[17]. The smaller the grain size, the more the grain boundaries, the higher the resistivity, the lower the RRR. We examined the microstructures of samples of different RRR by SEM. Fig. 5(a)-(c) are cross-sectional ion beam images of Cu stabilizer with RRR = 25, 48, and 87 respectively. Apparently, samples with smaller grain size have lower RRR. Due to the ion channeling effect, grain boundaries and twin boundaries are clearly seen in these ion beam images. The twin boundaries can be distinguished from grain boundaries by its straightness. It should be noted that Fig. 5(a) and (b) were taken from the Cu on the opposite side of the REBCO layer (the backside). With the identical electroplating conditions, the microstructure of the backside Cu should be identical to that of the frontside where the RRR measurements were taken. The average grain sizes were analyzed from Fig. 5 by counting number of intersections between grain boundaries and horizontal lines that are at different distances from the Cu/Ag interface. In these analyses, we ignored twin boundaries. Because twin boundaries have at least one order of magnitude lower resistivity than grain boundaries [15]. Table 2 lists the grain sizes measurement results, which indicates that RRR values are strongly correlated with the grain size. It is postulated that grain boundaries are mostly responsible for the resistivity in these samples. Another observation made from Fig. 5 is that grain size is significantly smaller near the Cu/Ag interface where the electroplating process started. The grain size gradually increased as the Cu grew thicker. This could partially be the reason why the average RRR for the 32 T magnet conductors is slightly higher than that for the 40 T magnet conductors.

**Table 2. Cu grain size measured from Fig. 5**

| Distance from Cu/Ag interface ($\mu m$) | Grain size ($\mu m$) | | |
|---|---|---|---|
| | RRR = 25 | RRR = 48 | RRR = 87 |
| 5 | 1.3 | 1.9 | 2.5 |
| 10 | 1.4 | 2.0 | 2.7 |
| 15 | 2.0 | 2.0 | 3.6 |
| Average | 1.5 | 2.0 | 2.8 |

### 3.4 Annealing Cu stabilizer

Thermal annealing of Cu reduces the density of dislocations, twin boundaries, and enlarge the grain size through the recovery process. So an experiment was designed to verify that the increase in RRR is concurrent with the increase of grain size by annealing, which would support the postulation that RRR in our samples was dominated by grain boundary resistivity. Table 3 lists RRR values of several samples before and after annealing of REBCO samples at 300 C for 30 minutes in argon. As expected, RRR increased considerably by the annealing for all the samples. Meanwhile the grain size was also significantly increased as shown by SEM images in Fig. 6 (for sample D in table 3). This concurrence of RRR increment and grain growth was indeed confirmed.

**Table 3. Effect of annealing at 300 C for 30 minutes in argon**

| Sample | RRR before | RRR after |
|--------|------------|-----------|
| A | 48 | 73 |
| B | 23 | 65 |
| C | 24 | 85 |
| D | 25 | 76 |

Annealing at 300 C will cause significant degradation in critical current of REBCO tapes [18]. To explore the possibility of improving Cu RRR without degrading REBCO tapes, we annealed a low RRR (RRR = 25) REBCO sample at moderate temperatures of 80 - 140 C for 2 hours. Annealing at these temperatures, as shown in Fig. 7(a), did not improve RRR much. Significant improvements were observed only at temperatures above 150 C (for 30 min.). Therefore, annealing REBCO is not a practical method to improve RRR without suffering a loss in critical current. Based on the 0.5 hour anneal data in Fig. 7(a), an Arrhenius plot of RRR increment ($\Delta$RRR) is fitted with a straight line, as shown in Fig. 7(b). An activation energy of 0.4 eV is obtained.

## 4. Discussions

*4.1 Influence of the Ag layer*

Due to the additional stabilizing effect of the 2 μm Ag layer, RRR was measured on the Cu/Ag composite instead of just Cu. But in the investigation of factors influencing RRR, we only studied impurity and microstructure of the Cu layer. So it is prudent to prove that RRR of Cu is approximately the same as RRR of Cu/Ag, given the relatively small thickness of the Ag layer. We prepared a Ag sample by chemically removing the Cu layer from the Cu/Ag composite film with the APS-100 Cu etchant. The RRR of the Ag layer was measured to be 15. Since the resistivity of Cu and Ag are very similar at room temperature [19], using a parallel circuit model, RRR of Cu/Ag composite can be written approximately as,

$$RRR_{Cu/Ag} = (RRR_{Cu}t_{Cu} + RRR_{Ag}t_{Ag})/(t_{Cu} + t_{Ag}) \quad (3)$$

Where *t* denotes the layer thickness. If the thickness of the Ag layer is much smaller than that of the Cu layer, $RRR_{Cu/Ag}$ is nearly the same as $RRR_{Cu}$. In our case of $t_{Ag}$ = 2 μm, $t_{Cu}$ = 20 or 50 μm, and $RRR_{Ag}$ = 15, $RRR_{Cu/Ag}$ is only slightly lower than $RRR_{Cu}$. For example, if $RRR_{Cu}$ = 50, the $RRR_{Cu/Ag}$ is 47 and 49 for 20 μm and 50 μm Cu respectively. Since the difference is small and comparable to the measurement error, we use the measured RRR of Cu/Ag and RRR of Cu interchangeably.

*4.2 Effect of Cu microstructure*

Equation (1) indicates that Cu resistivity increases with microstructural defects which include dislocations, grain boundaries and other defects. For instance, Cu resistivity increases significantly with cold work which introduce microstructural defects [10]. The effect of grain boundaries on Cu resistivity is well established [13]-[17]. The grain boundary resistivity $\rho_G$ can be related to the grain size *d* by a simple relationship [13],

$$\rho_G = \rho_0 + A/d \qquad (4)$$

Where $\rho_0$ is the resistivity other than from grain boundaries. $A = 7.22 \times 10^{-16}$ $\Omega\text{-m}^2$ is a constant obtained from fitting experimental data [15]. If $\rho_0$ is negligibly small at low temperatures, $\rho$ is dominated by the second term. Then Eq. (4) can be rewritten in term of RRR as,

$$RRR = K \cdot d \qquad (5)$$

Where $K = 23.8$ $\mu m^{-1}$. In such case, RRR is proportional to the grain size, and a grain size of 1 $\mu m$ corresponds to a RRR of 23.8. In Fig. 8, we plot the measured RRR versus the grain size obtained from table 3. The estimated measurement uncertainty in grain size is ± 0.5 $\mu m$. For comparison, the solid line is calculated by equation (4) assuming $\rho_0 = 8.6 \times 10^{-11}$ $\Omega\text{-m}$, which corresponds to $RRR_0 = 200$ without grain boundary resistivity. A straight dashed line by equation (6) are also plotted in the figure. The agreements between the experimental data and the calculated lines are reasonable, given the considerable uncertainties in grain size measurement of this work and ref. [17] where parameter A was obtained. The results in Fig. 8 rule out significant contribution from chemical impurities. Because a significant amount of impurity would make $\rho_0$ higher and RRR lower, which would make the calculated solid line further deviate from our experimental results. Without significant contribution from impurities, grain boundary resistivity became the dominant mechanism responsible for the observed RRR. Similar trend of RRR versus grain size of electroplated Cu films was reported in ref. [20].

It is well known that electroplated Cu film undergoes a self-annealing within a few tens of hours of deposition [21]-[23]. In the self-annealing process, the grain size grows at room temperature, and resistivity decreases by up to 20% [21]. In other words, the resistivity of the freshly electroplated Cu is 25% higher than that stored at room temperature for a few days. This 25% extra resistivity is due to microstructural defects which would remain the same at low temperatures. Therefore, it can be deduced that the RRR of freshly plated Cu is in the order of 5. After the self-annealing, the RRR of our samples increased to 30 - 60. Such a dramatic RRR enhancement highlights the importance of the self-annealing process. Any mechanism that hinders the self-annealing process would inevitably cause low RRR as will be discussed in the next section. After a few days of self-annealing, further increment of RRR becomes negligibly small. In fact, we measured a sample with RRR = 25 six months after the initial measurement, RRR remained the same.

Additional annealing at elevated temperatures can further enhance RRR as shown in Fig. 7(a). Similar annealing effect was reported in ref. [24]. Since we proved that RRR is dominated by grain boundary resistivity, the annealing effect must correspond to grain growth which was indeed observed by microscopy. The activation energy of 0.4 eV from Fig. 7(b) is in reasonable agreement with the activation energy of 0.3 – 0.9 eV obtained from Cu recovery experiments [25], [26]. This is another evidence that resistivity due to microstructural defects dominated RRR.

*4.3 Impurity in electroplated Cu*

Chemical impurity, especially oxygen, is usually the biggest contributor to residual resistivity in bulk Cu. During the electroplating process, however, the Cu film (cathode) has an accumulation of reduced hydrogen which inhibits high level of oxygen from forming in the electroplated film. This explains the relatively low oxygen content in our samples measured by SIMS. To make Cu films smooth and bright, sometimes additives, which often contains S, P, and Cl, are added in the plating bath [27]. In such case, S, P, and Cl are often detected in Cu, which can result in low RRR. Since there were no additives in the plating bath for the SuperPower conductors, the concentrations of S and P were indeed negligibly low. Cl concentration, however, was surprisingly high especially in low RRR samples, even though no Cl was intentionally added in the bath. Moreover, it is unexpected that Cl concentration is strongly correlated with RRR. Because Cl dissolved in Cu has not been reported to be particularly detrimental to conductivity, and we proved that microstructural defects are mostly responsible for RRR of our samples. Interestingly, it was previously reported that high concentration of Cl in electroplated Cu hinders the grain growth in the self-annealing process [28]. This is consistent with our observation that samples with higher Cl have smaller grain sizes. Therefore, we postulate that high concentration of Cl deterred grain growth during self-annealing. This resulted in small grain size in high Cl samples. The small grain size in-turn resulted in low RRR, since grain boundary resistivity is mostly responsible for RRR. This explains why high Cl samples exhibit low RRR.

## 5. Conclusions

The RRR of Cu stabilizers of over 130 REBCO samples were measured. Their typical RRR value was 50. Electron microscopy revealed that Cu grain size increased with increasing RRR, which is in agreement with the relationship between Cu resistivity and grain size in the literature. This proved that grain boundary resistivity is mostly responsible for the RRR in our samples. Annealing at elevated temperatures resulted in grain growth with an activation energy of 0.4 eV. As a result, RRR of samples annealed at 300 C is considerably higher. Chemical impurities O, Cl, S, P, Fe and other elements in Cu were measured by SIMS and ICP-MS. Cl concentration was strongly correlated with RRR. This is explained by that relatively high concentration of Cl deterred Cu grain growth during the self-annealing at room temperature, and smaller grain size resulted in low RRR due to grain boundary resistivity.

**Acknowledgement**

Dr. Hang Dong Lee and Dr. Steven Liu of Eurofins / EAG Laboratories are acknowledged for their contribution in SIMS analyses. This work was performed at the National High Magnetic Field Laboratory, which is supported by National Science Foundation Cooperative Agreement No. DMR-1644779, DMR-1839796, DMR- 2131790, and the State of Florida.

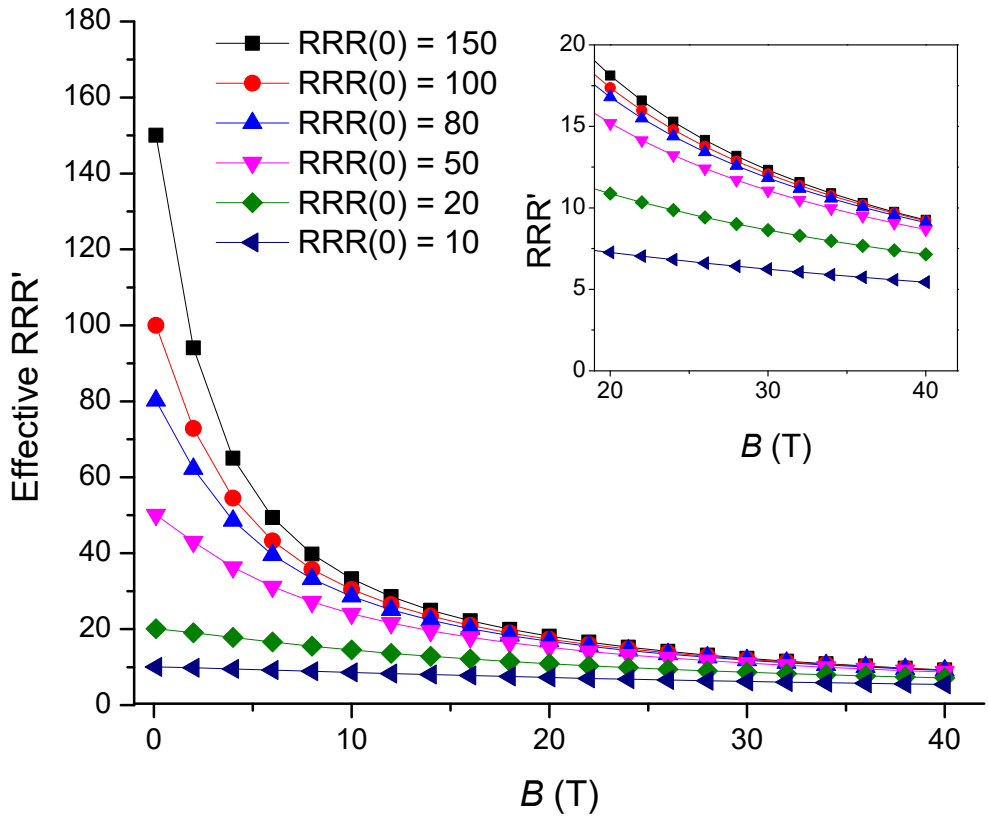

Fig. 1. The effect of magnetoresistance on Cu conductivity. Effective RRR as a function of magnetic field RRR (B) for Cu of different zero-field RRR which is denoted as RRR (0). The inset is a close-up for magnetic fields above 20 T.

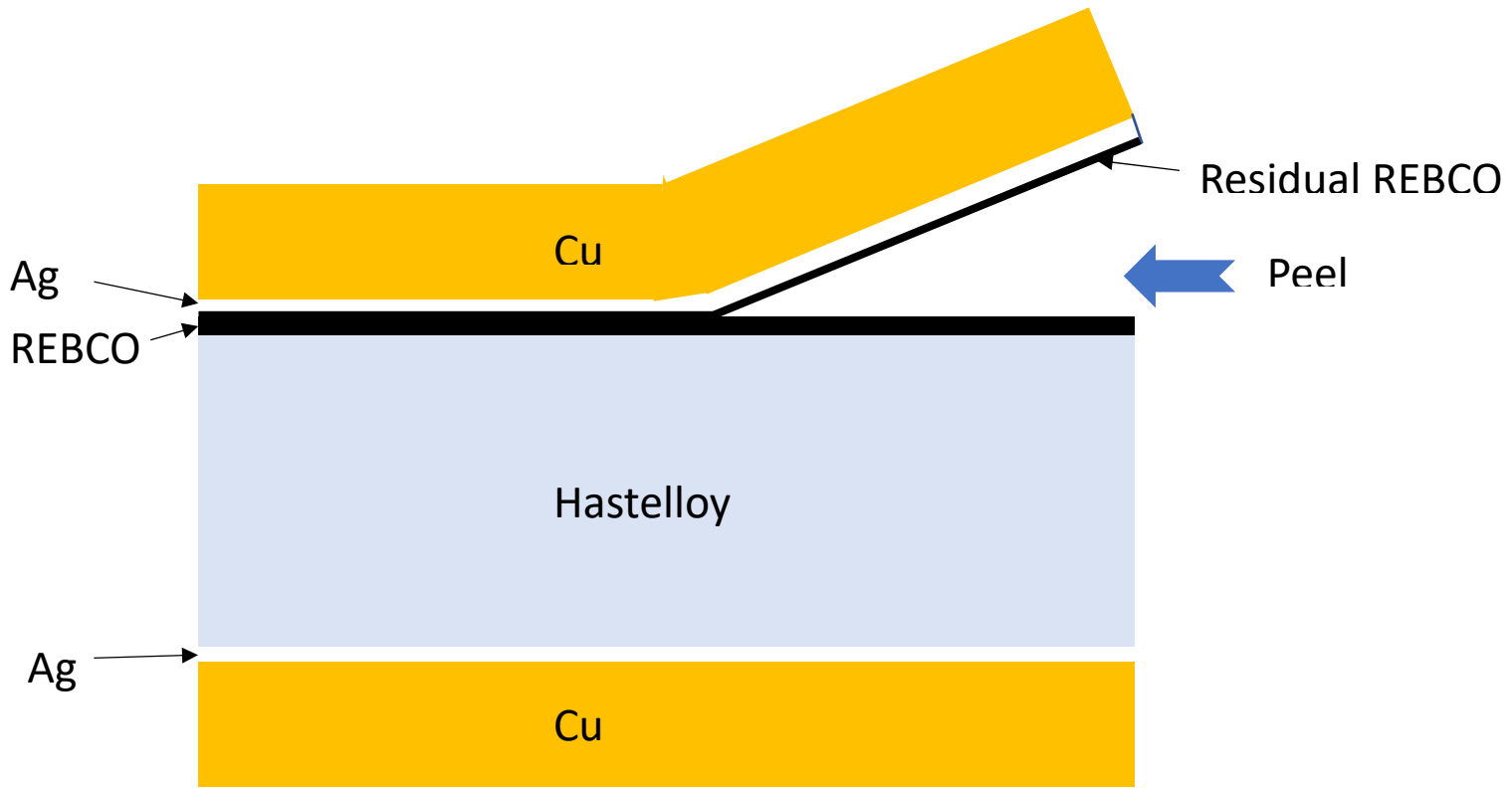

Fig. 2. A schematic of the layer structure of REBCO coated conductor where Cu/Ag layer above the REBCO layer is peeled off for RRR tests.

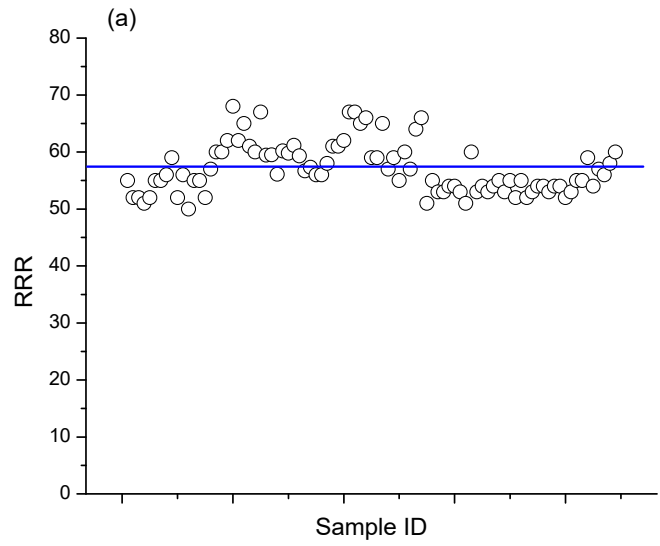

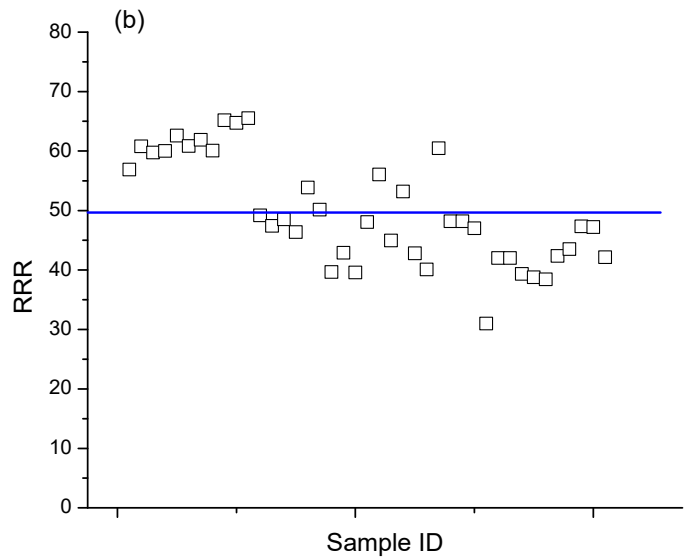

Fig. 3. Measured RRR of Cu stabilizer of REBCO for (a) 32 T magnet project conductors with 50 μm Cu (total 89 data), (b) 40 T magnet project conductors with 20 μm Cu (total 42 data). The solid horizontal lines represent the average RRR values.

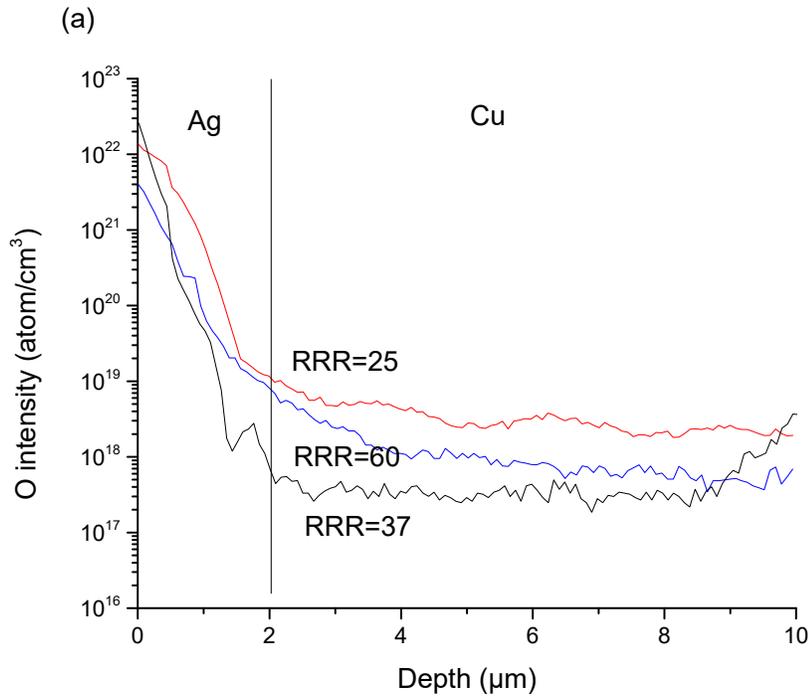

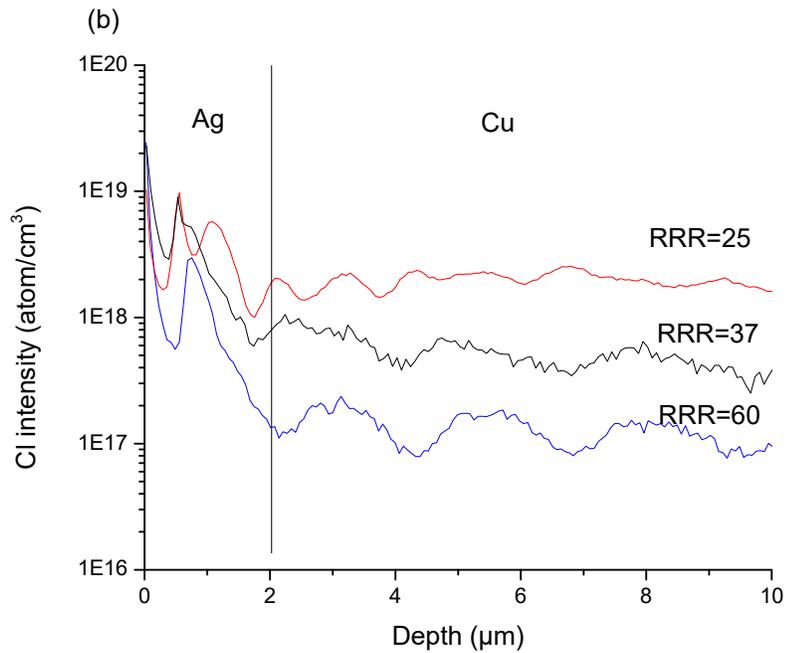

Fig. 4. SIMS depth profiles of (a) O in Cu stabilizer of different RRR, (b) Cl in Cu stabilizer of different RRR. The profiling started from the Ag layer. The initial 2 µm is the Ag layer which is known to permeable to O.

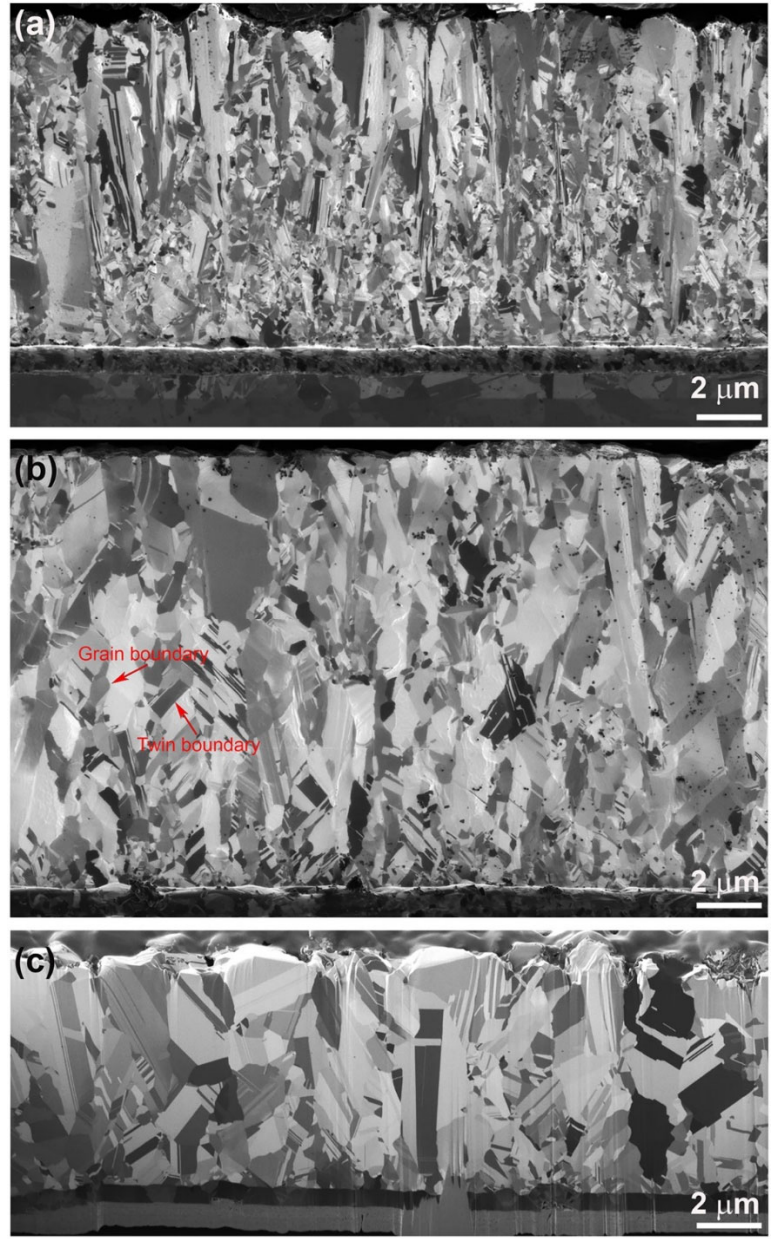

Fig. 5. SEM ion beam images of Cu stabilizers (a) RRR = 25 (b) RRR = 48, (c) RRR = 87

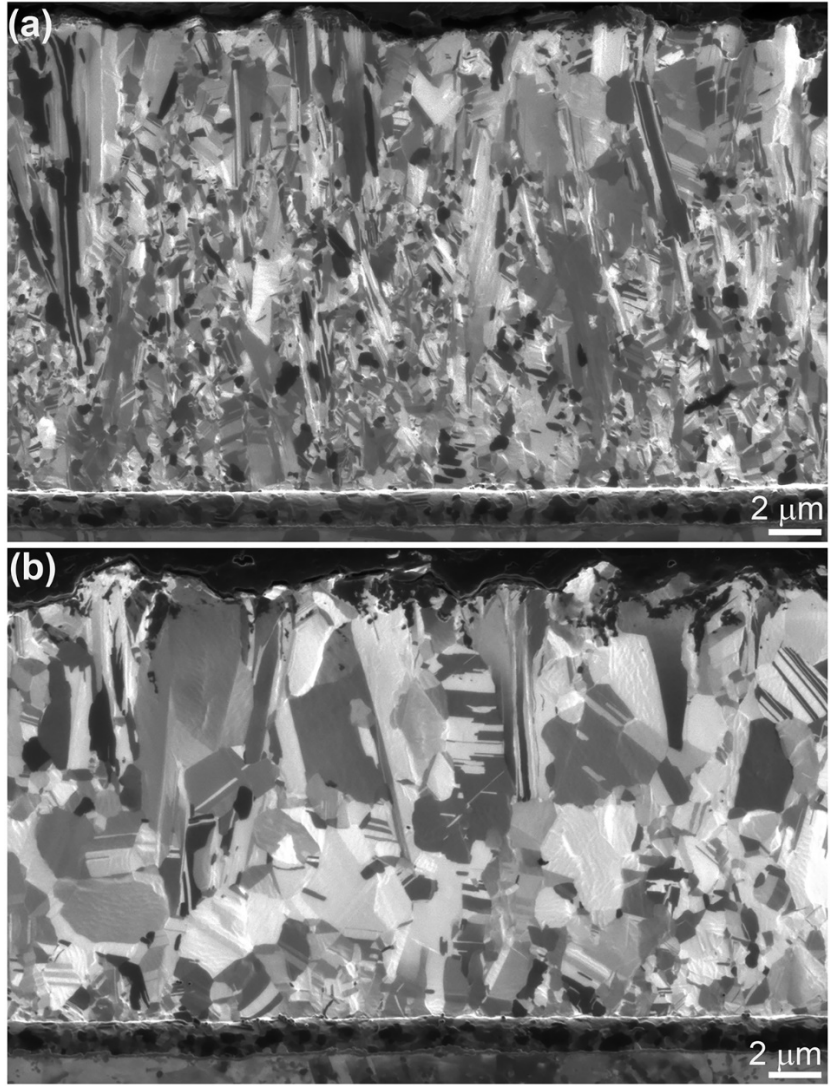

Fig. 6 SEM ion beam images of a sample of RRR = 25 before (a) and after (b) annealing at 300 C for 30 minutes. After annealing RRR was measured to be 76 (sample D in table 3).

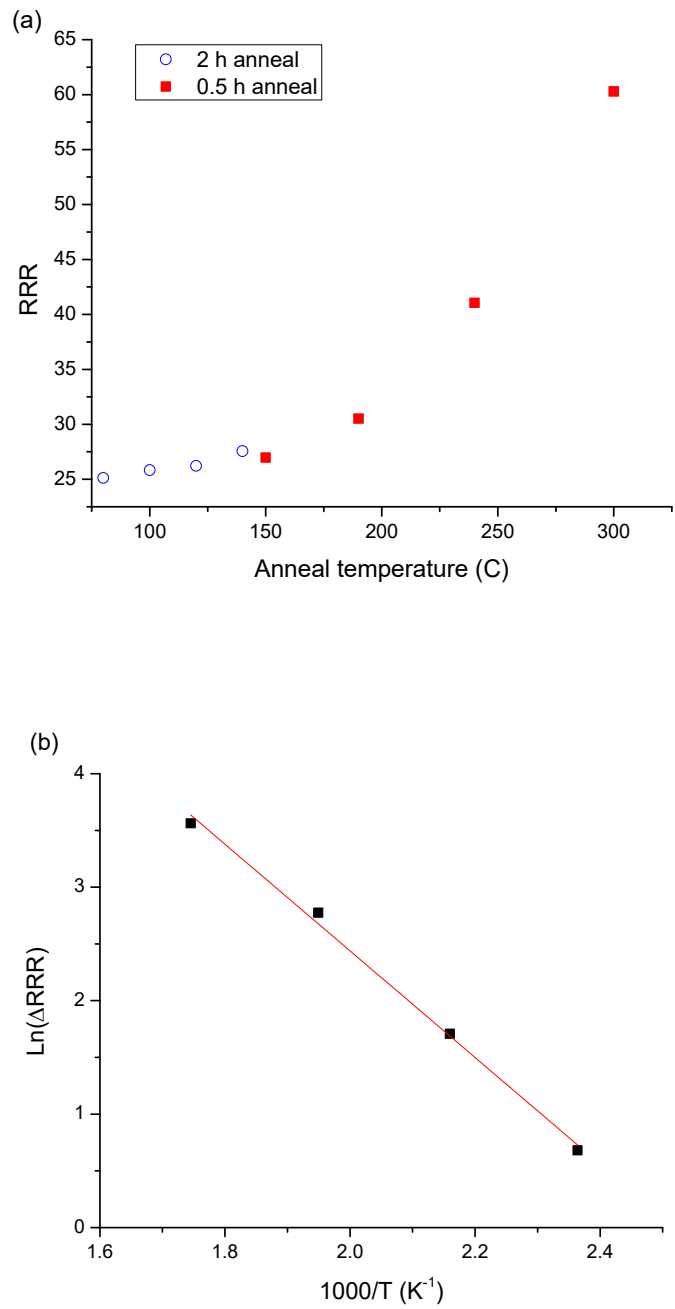

Fig. 7. Effect of annealing on RRR. RRR was 25 before annealing. (a) RRR vs. annealing temperature. (b) Arrhenius plot of the 2 hour anneal data in (a). An activation energy of 0.4 eV is obtained from the linear fit.

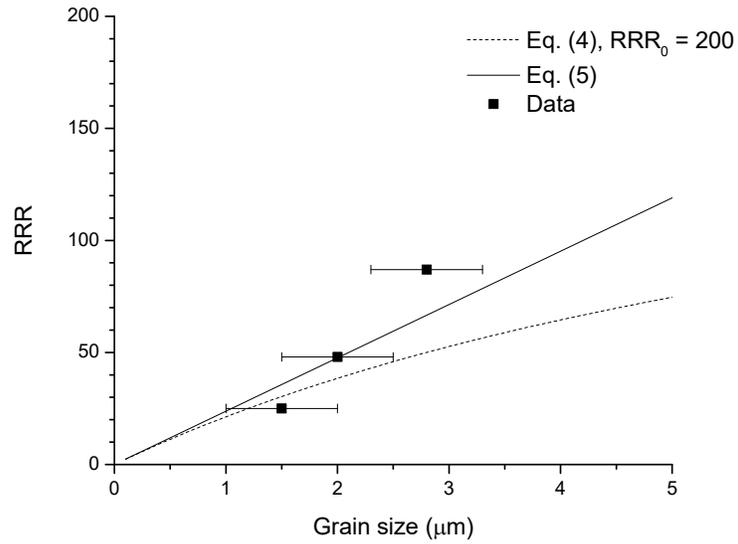

Fig. 8. RRR versus Cu grain size compared with equations (5) and (4) where an $RRR_0$ (without grain boundary) of 200 is assumed. The estimated uncertainty in the measured gain size is 0.5 μm. The dashed and solid lines are the prediction by equation (4) and (5) respectively.